# Track-Dependent Links between Tropical Cyclones and Extratropical Predictability in Physical and AI Models

**Gan Zhang**[1*]

[1] Department of Climate, Meteorology & Atmospheric Sciences, University of Illinois Urbana-Champaign, Urbana, Illinois

*Corresponding Author: Gan Zhang (gzhang13@illinois.edu)

## Key Findings

- Physical and AI-hybrid models can make comparable forecasts of extratropical large-scale flow up to two weeks after TC genesis.
- TC associations with extratropical predictability are track-dependent, and significant skill changes rarely occur across all metrics.
- While poleward-moving TCs expectedly affect US-European forecasts, experiments also suggest significant impacts from westward-moving TCs.

## Abstract

Global medium-range weather forecasts suffer occasional failures ("busts"), often linked to tropical cyclones (TCs). We investigate TC influences on extratropical predictability by comparing forecasts from a physics-based model (ECMWF-IFS) and an AI-hybrid model (Google-NGCM) initialized near TC genesis. Analyzing 108 out-of-sample Northern Hemisphere cases reveals similar extratropical error growth patterns and comparable performance between the models. This suggests that the NGCM is capable of predicting the bulk upscale effects of tropical convection without directly representing convective processes. Leveraging the NGCM's computational efficiency, we compare forecasts initialized with and without TC genesis to isolate track-dependent forecast impacts. For Week-2 extratropical forecasts, TC impacts are highly time-, metric-, and track-dependent. The analysis confirms that some poleward-moving TCs degrade Week-2 US and European forecasts and suggests significant impacts from westward-moving TCs. The findings highlight the utility of the AI-hybrid model in predictability research and complex tropical-extratropical teleconnections that warrant future research.

## Plain Language Summary

Investigations of hurricane forecasts usually focus the direct local impacts of the storm itself; however, hurricanes can affect weather patterns thousands of miles away by altering the large-scale flow. This study uses both traditional and artificial intelligence (AI) weather models to investigate how these distant storms impact 14-day forecasts. We demonstrate that an AI model is highly capable of predicting the large-scale outcome of these complex, long-distance interactions, operating on par with traditional physics-based models. Crucially, we reveal a stark geographic divide between storms on how they affect midlatitude predictions. While Atlantic hurricanes generally make European weather harder to predict 1-2 weeks in advance, the impact on the United States depends more on a storm's path. Surprisingly, certain Pacific and Atlantic storms are associated with better US weather predictions. These insights help forecasters know when to confidently trust long-range weather outlooks and highlight the growing value of AI in global weather prediction.

# 1. Introduction

The numerical weather prediction (NWP) has been one of the great scientific achievements of the last century (Bauer et al., 2015). Predictions of large-scale weather patterns are now routinely skillful out to two weeks or more, assisting weather-sensitive applications. Despite this progress, these predictions are prone to occasional but significant failures, known as "forecast busts", where the forecast rapidly deviates from the observed atmospheric evolution (Rodwell et al., 2013). Understanding the precursors and dynamics driving these busts is crucial for improving forecast reliability and potentially extending the horizon of useful weather predictions.

Research on forecast busts, often focusing on European Week-1 predictions, frequently links these errors to upstream meteorological precursors. Rodwell et al. (2013) suggested that springtime busts for Week-1 prediction are often associated with North American mesoscale convective systems (MCSs), which trigger Rossby wave trains across the North Atlantic. Hauser et al. (2026) noted that forecast busts occur more frequently under large-scale cyclonic regimes in the North Atlantic-European sector. The sources of forecast degradation shift seasonally (Hauser et al., 2026; Lillo & Parsons, 2017). For instance, Lillo & Parsons (2017) identified a prominent seasonal peak of European forecast busts in September and October. This period coincides with the Atlantic hurricane season peak, pointing to a strong, though not exclusive, link between tropical cyclone (TC) activity and Week-1 forecast busts.

The basic physical mechanism linking recurving TCs to downstream forecast error is conceptually well-understood and supported by modeling studies (Keller et al., 2019). A poleward-moving, recurving TC can interact with the mid-latitude jet and undergo extratropical transition (ET; Evans et al., 2017). The TC trajectory and subsequent flow evolution are sensitive to the phasing of TC tracks and midlatitude waves (Riboldi et al., 2019; Riemer & Jones, 2014). The diabatic heating and potential vorticity (PV) anomalies associated with the recurving cyclone can perturb the midlatitude jets (Archambault et al., 2015; Evans et al., 2017), radiating Rossby wave trains that propagate downstream (Archambault et al., 2013; Keller et al., 2019; Quinting & Jones, 2016). In some cases, amplifying ridges and troughs alter the large-scale flow pattern thousands of kilometers away, leading to high-impact weather (e.g., Pohorsky et al., 2019; Riboldi et al., 2019) or a rapid decrease in forecast skill (Aiyyer, 2015; Quinting & Jones, 2016; Harr & Archambault, 2017). While the downstream impact is highly dependent on the TC genesis location and subsequent track (Grams et al., 2015; Grams & Archambault, 2016; Keller et al., 2019), TC intensity has been shown to have relatively small impact on downstream wave characteristics (Riemer & Jones, 2010; Archambault et al., 2013; Riboldi et al., 2019).

Increasing evidence suggests potential impacts of TCs on extratropical flow may occur before the extratropical transition. Interactions with midlatitude systems can export tropical moisture from the vicinity of TCs to higher latitudes, resulting in predecessor rain events (Cordeira et al., 2013; Galarneau et al., 2010) that affect the downstream flow. Even when residing in the tropics, TCs can interact with equatorward intruding Rossby waves and the associated low-level anticyclonic circulation (Zhang et al., 2016, 2017), which can transport moisture to the midlatitudes (Liu et al., 2026). Furthermore, Sinclair (2025) showed that observed non-recurving TCs can excite or amplify of midlatitude Rossby waves.

Understanding TC-induced forecast busts has traditionally been challenging. Past studies of the impacts of recurving TCs on downstream weather predictability often rely on NWP case simulations (Riemer & Jones, 2014; Grams & Archambault, 2016; Grams et al., 2015), which are often limited by the computational expense of running ensemble simulations. Past studies analyzed archived operational forecasts (Aiyyer, 2015; Anwender et al., 2008) and model reforecasts (Harr & Archambault, 2017; Quinting & Jones, 2016). However, they were limited to small sample sizes due to frequent updates of operational NWP models or the sparse initialization time of reforecasts. Finally, past studies of forecast busts mostly focus on Week-1 forecasts initialized around TC recurving, leaving Week-2 forecast busts and potential TC impacts understudied.

Artificial Intelligence (AI) is transforming global weather forecasting (e.g., Bi et al., 2023; Lam et al., 2023; Lang et al., 2024; Price et al., 2023; Kochkov et al., 2024; Bonev et al., 2025; Lang et al., 2026). These new models can produce forecasts orders of magnitude faster than traditional NWP models. Leveraging AI models' computational efficiency to explore vastly more historical forecasts over longer lead times alleviates past limitations regarding small sample sizes. As AI models enters operation and become popular, it is also valuable to compare the behaviors of physical and AI models in highly challenging forecast scenarios. Since current AI models do not directly represent convective processes involved with TC internal dynamics or extratropical transition, a key question arises: Can an AI model accurately simulate the large-scale flow evolution after TC genesis as skillfully as a physical model?

The Neural General Circulation Model (NGCM) (Kochkov et al., 2024) is uniquely positioned for answering the above question. Combining a numerical dynamical core with machine-learned subgrid parameterizations, the NGCM has delivered skillful two-week weather prediction (Kochkov et al., 2024) and seasonal prediction of TC activity (Zhang et al., 2025). Crucially, its hybrid architecture represents the large-scale dynamics as physical models and enables testing whether explicitly resolving convective processes is strictly necessary to capture TC-induced extratropical flow responses. With a focus of the Northern Hemisphere TCs, we compare an operational physical model (Integrated Forecasting System; IFS) and an AI-hybrid model (NGCM) to investigate:

1. Does the AI-hybrid NGCM capture the impacts of TCs on extratropical forecasts as well as the physical IFS?
2. How may specific TC track archetypes modulate extratropical forecast skill extending into Week-2?

By pivoting away from the TC recurvature and analyzing ensemble forecasts initialized near TC genesis, we characterize the connection between TCs and extratropical predictability loss in a broader context. The comparison between physical and AI forecasts offers insights relevant to both operational forecasting and ongoing model development.

# 2. Data and Methods

## 2.1 Tropical Cyclone Tracks and Clustering

Historical atmospheric states and TC tracks of 1979–2022 were sourced from the European Centre for Medium-Range Weather Forecasts (ECMWF) reanalysis version 5 (ERA5) dataset (Hersbach et al., 2020). Specifically, the ERA5 tracks were identified by Han & Ullrich (2025) using an algorithm that detects and tracks cyclone features based on sea-level pressure minima, wind speed thresholds, and warm-core characteristics. Even though TCs extracted from reanalyses suffer from biases (e.g., intensity), their tracks are generally consistent with the best-track observational data, especially for stronger storms (Han & Ullrich, 2025; Hodges et al., 2017). This study uses ERA5 tracks strictly to determine TC genesis time for forecast initialization. Using a single, globally consistent dataset avoids potential inhomogeneities and artificial boundaries between different operational agencies, helping consistently assess TC-associated forecast skill changes across basins.

To group TC tracks that share similar spatial characteristics, we apply a trajectory clustering algorithm (TRACLUS) (Lee et al., 2007). The algorithm segments each trajectory based on its spatial geometric properties, effectively creating a standardized vector representation. We then apply K-means algorithm to these trajectory representations to group similar tracks for the North Atlantic (NA), Central and Eastern Pacific (CEP), and Western Pacific (WP) basins. The resulting clusters were labeled (a) through (d), arranged geographically from west to east. The cluster number choice (k=4) follows previous studies of the North Atlantic (Kossin et al., 2010; Camargo et al., 2021) and balances representativeness and simplicity. Additional discussion of the clustering robustness is available in Supplementary Materials (Figures S2-S5).

## 2.2 Forecast Models

This study uses two global weather forecast models with distinct architectures. We acknowledge that the findings may not fully generalize to other physical and AI models.

- The IFS is the operational global NWP model developed by ECMWF. This spectral model solves the primitive equations of atmospheric motion, incorporating a sophisticated suite of physical parameterizations to represent processes like convection, radiation, and turbulence. This study uses a set of IFS forecasts sourced from the THORPEX Interactive Grand Global Ensemble (TIGGE) archive (Bougeault et al., 2010) covering 2020–2022, approximately corresponding to IFS Cycle-47 version. During this period, the IFS ensemble consisted of 50 members, generated using an Ensemble of Data Assimilations (EDA) and singular vector perturbations, running at a horizontal resolution of ~18 km.

- The AI-based NGCM employs a numerical dynamical core to resolve large-scale atmospheric dynamics but replaces traditional sub-grid physics parameterizations with components learned directly from ERA5 data (Kochkov et al., 2024). While purely data-driven models often suffer from excessive spatial smoothing, the NGCM architecture resolve features more comparably to traditional NWP models of similar grid spacing. We primarily analyze the ensemble forecasts from the 1.4-degree stochastic NGCM. The stochastic NGCM injects spatially and temporally correlated Gaussian noise into its encoder and forward step as additional input to provide randomness for ensemble forecasts. Its training used the Continuous Ranked Probability Score (CRPS) to balance accuracy with ensemble spread and ingested ERA5 data of 1979–2019. For completeness, the Supplementary Materials include results from the deterministic NGCM, which lacks probabilistic inputs and mainly optimizes for accuracy and sharpness.

Crucially, the AI-based parameterizations within the NGCM do not intrinsically represent the convective processes or internal dynamics involved in TC development and extratropical transition. A key focus of the analysis is thus whether the NGCM can predict the large-scale extratropical flow responses despite its fundamental architectural difference from the physics-based IFS.

## 2.3 Experiment Design and Evaluation

The evaluation between models and across track clusters follows a three-step procedure (Figure S1). Section 3.1 first analyzes two individual cases to demonstrate relevant dynamic processes. To enable direct comparisons, we coarsen the ERA5 and IFS data to the NGCM grid and compute 200-hPa horizontal wave activity fluxes (WAF; Takaya & Nakamura, 2001). The WAFs are parallel to Rossby wave group velocity and trace how TC-amplified wave energy propagates into the midlatitudes and may drive forecast error growth. Section 3.2 analyzes the 108 forecast cases during 2020–2022 to validate the out-of-sample skill of NGCM forecasts against the IFS. After establishing this out-of-sample performance baseline, Sections 3.3 expands the NGCM evaluation to 359 total historical TC cases to evaluate relative skill differences across individual track clusters. This expanded analysis encompasses dates both within and outside the NGCM's 1979–2019 training period; however, focusing on the relative skill differences across track clusters helps mitigates potential in-sample biases and ensures a robust comparative assessment.

To generate the NGCM historical cases, we select the 30 most recent TC genesis events for 11 track clusters (Table S1). Since the CEP Cluster (a) contains only 29 cases during 1979-2022, we work with all the 29 available cases. To keep the computation and data volume manageable, we perform 20-member ensemble prediction with the NGCM. All ensemble members are initialized from the same ERA5 data corresponding to 0000 UTC on the day closest to the TC genesis, defined as the first time reaching 34-knot intensity. The stochastic NGCM generates ensemble spread through injecting space-time correlated Gaussian noise into the encoder and physics modules of the NGCM (Kochkov et al., 2024). To evaluate the NGCM's out-of-sample skill (Section 3.2), we focus on the 2020–2022 subset (N=108) of the NGCM forecasts and compared them with the corresponding IFS ensemble initialized with explicit initial condition perturbations. For a comparable ensemble size to the stochastic NGCM setup and practical reasons, we analyze the first 20 ensemble members of these IFS forecasts. As the IFS ensemble uses stochastic perturbations, the choice of 20 members remains arbitrary independent of which specific ensemble members are chosen. Different from previous studies that initialize experiments near TC recurvature, the initialization strategy here captures potential TC impacts even before they reach the midlatitude. Lastly, to establish a reference baseline, we conduct NGCM forecasts initialized on all 360 days during July–November of 2020–2022 free of a TC genesis in WP, CEP, and NA.

Following Rodwell et al. (2013), our skill analysis focuses on the 500-hPa geopotential height (Z500), a key indicator of the large-scale mid-troposphere flow pattern. The consistency between the ensemble means and the ERA5 is evaluated using fair CRPS (Ferro, 2014), Root Mean Square Error (RMSE), Anomaly Correlation Coefficient (ACC), and Mean Absolute Error (MAE). Since the distance-based metrics used in AI model training can inadvertently reward spatially smoothed predictions, we emphasize that high Z500 skill scores do not guarantee the physical realism of AI models (Bröcker et al., 2026; Casati et al., 2026), particularly at smaller scales or near the surface for TCs (DeMaria et al., 2025; Pasche et al., 2025). These Z500 metrics are utilized here strictly to evaluate the forecasts of the large-scale circulation, which is essential for establishing a baseline of model and predictability characteristics.

## 3. Results

### 3.1 Case Study of Forecast Error Growth

To illustrate the synoptic-scale evolution and error growth mechanisms related to TCs, we first examine the Z500 forecasts for Hurricane Larry (2021) and Hurricane Laura (2020) (Figure 1). These two TCs precede substantial error growth of operational North Atlantic-European forecasts. Before recurving, Hurricane Larry moved northwestward and merged with a tropical moisture plume associated with the residuals of Hurricane Ida (2021) (Figures 1a-b). Wave activity fluxes suggest Hurricane Larry, while still residing in the tropics, contributed to wave development in the East Atlantic (Day 5). After recurving, Hurricane Larry interacts with a midlatitude trough and amplifies an atmospheric block pattern, contributing to substantial forecast errors over the Atlantic (Day 11). In comparison, Hurricane Laura predominantly follows a westward track before landfall. The storm-associated moisture plume is redirected poleward and eastward by the large-scale subtropical high (not shown), merging with the residuals of Hurricane Marco (2020) and extending towards Europe (Figure 1g). Around landfall time, wave activity flux suggests Hurricane Laura strengthens the trough development near the US East coast. Furthermore, the residuals of Hurricane Laura feeds moisture to developing troughs near the US east coast, contributing to a wave train that resulted in substantial forecast errors over the North Atlantic (Figure 1h-i).

Although Figure 1 does not definitively diagnose the exact physical mechanisms, the large-scale error patterns are consistent with previously documented pathways. We speculate that the error growth in the Hurricane Larry case reflect a combination of interactions between equatorward-intruding trough (Figure 1a) and the low-latitude TC (Sinclair, 2025) and direct midlatitude interactions associated with the extratropical transition (Grams & Archambault, 2016). The error growth in Hurricane Laura case suggests potential influences from moisture transport resembling predecessor rain events (Bosart et al., 2012; Galarneau et al., 2010), and subsequent convective triggering of Rossby waves (Rodwell et al. 2013). Formally confirming these exact processes would require targeted analyses beyond the scope of this contextual review.

The error growth patterns in the IFS forecasts and the NGCM forecasts are highly consistent during the complex flow evolution. This consistency during complex flow evolution suggests that the NGCM's error growth and predictability limits closely mirror those of a high-resolution physical model. Since both hurricane cases are out-of-sample for the NGCM, the consistency and comparable skill here lend confidence in the NGCM's capability of simulating complex TC-impacted flows despite its coarser nominal grid spacing.

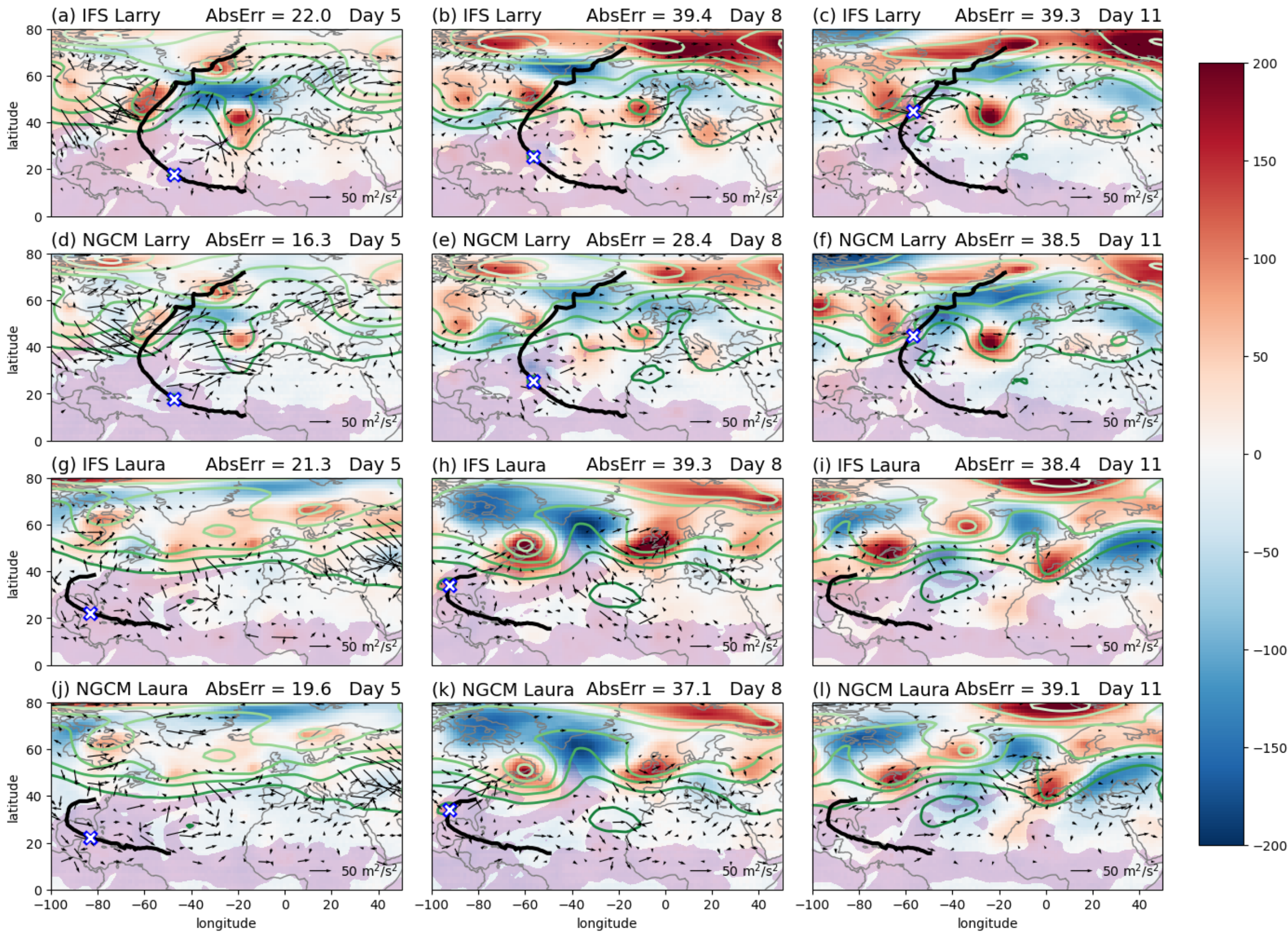

*Figure 1: ERA5 Z500 (contours, unit: m) and forecast errors (red-blue shading, unit: m) for Hurricane Larry (a-f; initialized on 2021-08-31) and Hurricane Laura (g-l; initialized on 2020-08-20). The ERA5 storm location is denoted by blue-edged white crosses, with the full track in black. Purple shading highlight ERA5 total column water vapor>40 mm. Forecasts are shown for Days 5, 8, and 11, for the IFS 20-member mean (a-c, g-i) and NGCM 20-member mean (d-f, j-l). Absolute errors (AbsErr) in the subplot titles are domain-averaged. The vectors show the 200-hPa wave activity fluxes, with low values masked out for clarity.*

## 3.2 Aggregated Forecast Skill of IFS and NGCM

To establish that out-of-sample NGCM forecasts and IFS forecasts perform comparably in TC-impacted cases, Figure 2 shows the time-latitude evolution of the forecast skill for 108 forecasts initialized near TC genesis in 2020–2022 (Table S1). Overall, the forecast skill gradually decays as the forecast lead increases. The decay accelerates in Week 2 and becomes more pronounced in the extratropics. Subsampling the IFS ensemble weakly affects the fair CRPS, but increases RMSE, and decreases ACCs moderately (Figure 2a-c), suggesting a larger ensemble size helps mitigate forecast errors. Additional spatial details of the IFS skill are available in Figure S6.

Compared to the subsampled 20-member IFS forecasts, the NGCM forecasts show regional strengths and weaknesses (Figure 2d-f). Specifically, while the NGCM forecasts show a skill deficit in the tropics at early lead times (higher CRPS and RMSE, alongside lower ACC), the skill differences between the two models become largely statistically insignificant in the extratropics, particularly after Day 4. Furthermore, in the extratropical regions where statistically significant differences are detected after Day 4, the NGCM generally demonstrates superior skill. This finding does not necessarily extend to other operationally critical variables, and many details also depend on forecast regions (Figure S7-S8). Nevertheless, establishing that the NGCM and the IFS perform comparably in predicting the extratropical Z500 in these out-of-sample forecast cases provides critical validation. This justifies our use of the expanded NGCM dataset to investigate how TC track archetypes modulate extratropical predictability differently.

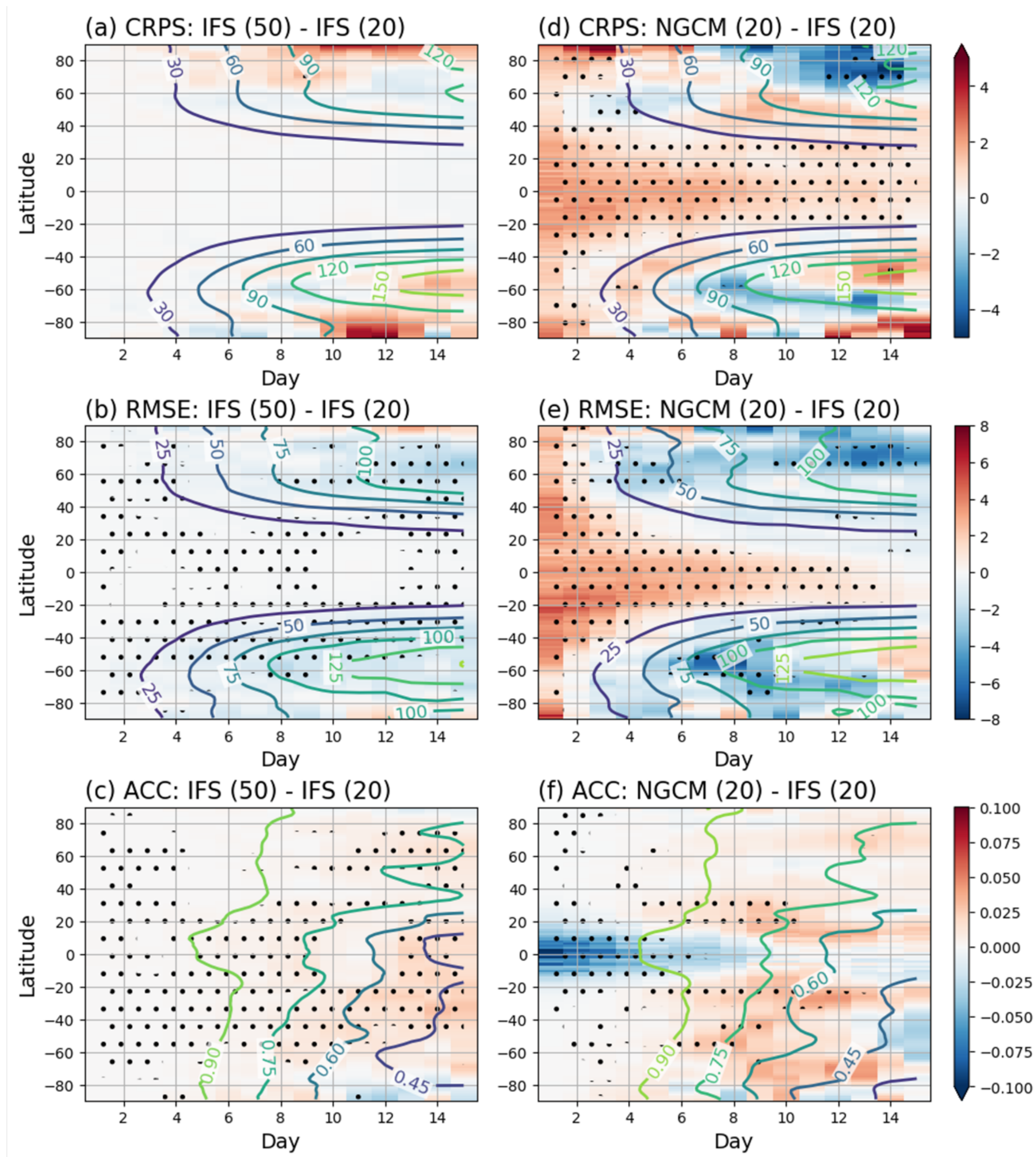

*Figure 2: Zonal average skill metrics for out-of-sample forecast cases in July-November of 2020-2022. Color shading shows differences for fair CRPS (a, d), RMSE (b, e), and ACC (c, f). (a-c): IFS 50-member minus 20-member subsets. (d-f): NGCM 20-member minus IFS 20-member subsets. ACCs and RMSE are computed using the ensemble mean. Stippling highlights 95% confidence level estimated with bootstrapping with replacement and the false discovery rate (Benjamini & Hochberg, 1995). Contours show the skill of the 20-member IFS forecasts.*

### 3.3 Characteristics and Impacts of Track Clusters

To investigate how TC tracks may affect extratropical forecast, we apply the objective clustering algorithm (Section 2.1) to identify track archetypes in each basin (Figure 3). The track clusters are relatively well separated and show differences in genesis locations and subsequent development (Figures S2-S5). For instance, westward-moving tracks are mostly limited to WP cluster (a), CEP cluster (b)-(c), and NA cluster (c), while poleward tracks dominate the rest clusters. Except for the NA Clusters (a) and (c), the tracks strongly affect the likelihood of extratropical transition, which were extensively studied for its downstream midlatitude impacts (e.g., Keller et al. 2019). Figure 3 also highlights that the NA Clusters (a)-(c) frequently recurve into the equatorward flank of the midlatitude jet entrance, while the CEP Clusters (b) and (c) generally move towards the weak subtropical jet and dissipate. Such contrasts likely affect the impacts of TC track clusters on the extratropical flow and predictability.

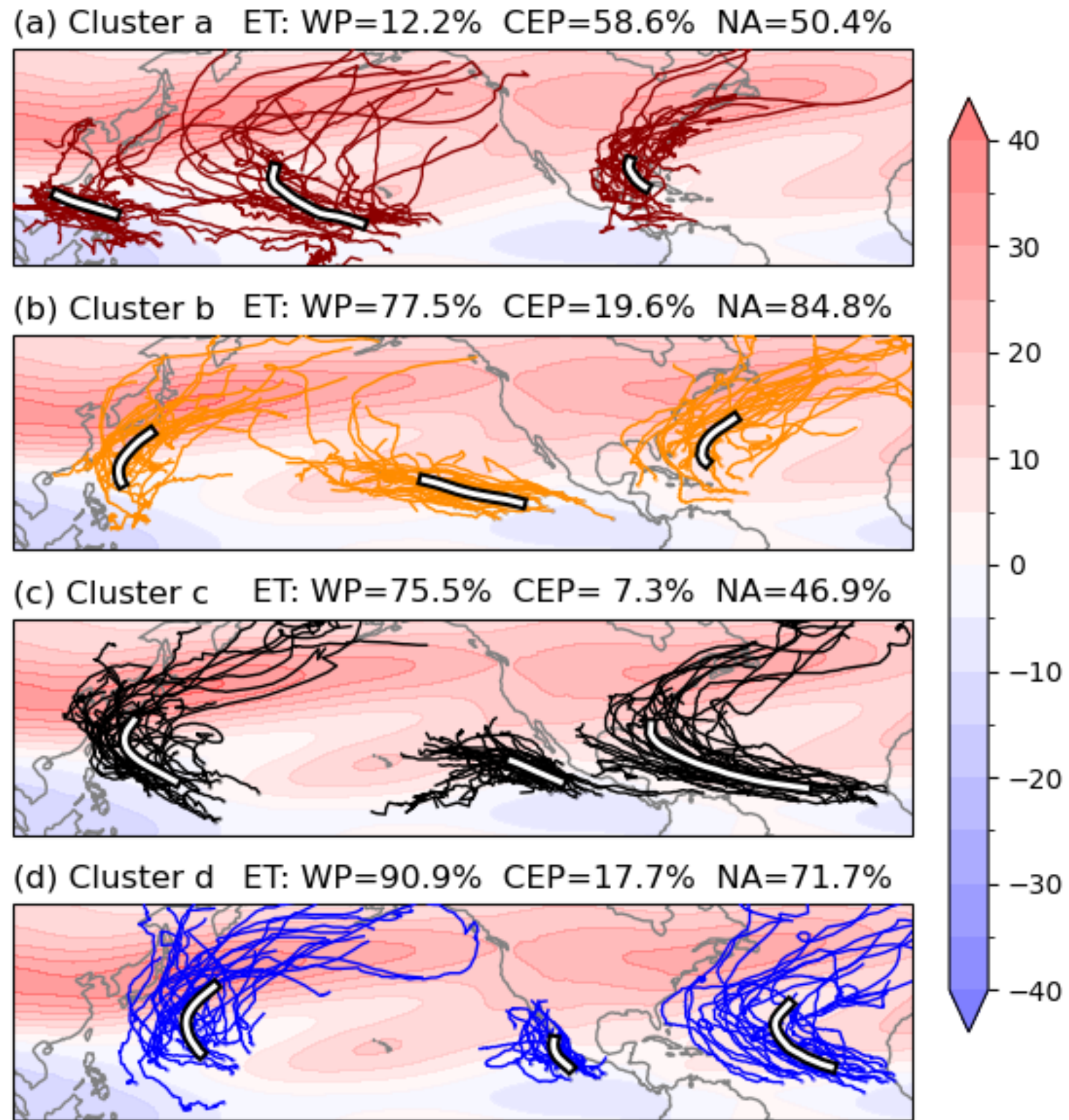

*Figure 3: TC track clusters (a-d) for NA, CEP, and WP basins (arranged geographically from west to east). Each panel displays 30 sampled tracks (thin lines) and cluster means (thickened while lines). Color shading shows the mean 200-hPa ERA5 zonal wind (unit: m $s^{-1}$) of June-November of 1990-2019. Extratropical transition percentages are denoted at the top right of each panel.*

We next examine the extratropical prediction skill associated with each track cluster using the extended NGCM forecasts (Section 2.3). The analysis here evaluates the NGCM across 359 forecast cases, drawing on samples from both within and outside the NGCM training period. By comparing the relative skill across track clusters, we mitigate the impact of potential in-sample performance biases. The following discussion focuses on CRPS, and additional results of ACC and RMSE are available in Figures S9-S10.

Comparing forecasts initialized with and without TC genesis suggests that certain track clusters degrade extratropical forecasts (Figure 4). During the first 3-4 forecast days, positive CRPS differences indicate larger initial errors for TC cases, possibly reflecting complex TC-genesis environments and minor differences in seasonal sampling compared to the non-TC baseline. Beyond Week 1, the impacts become more track-dependent. For instance, the westward-moving WP Cluster (a) degrades US forecasts in Week 2 (Figure 4i), while the NA Cluster (a) and (d) and WP Cluster (a) degrade European forecasts around Day 13 (Figures 4a, 4d, and 4i). The significance of skill changes also depends on the skill metric. For example, while ACCs suggest significant skill degradation associated with the NA Cluster (b) throughout Week 2 (Figure S9), RMSE (Figure S10) and CPRS (Figure 4b) suggest skill changes are generally insignificant.

Perhaps surprisingly, westward-moving TCs can accompany significant, yet contrasting, variations in midlatitude prediction skill despite rarely recurving. Specifically, WP Cluster (a) is the only cluster that degrades both US and European CRPS (Figure 4i). Preliminary analyses (not shown) suggest that these TCs are steered by a westward-extended Northwestern Pacific subtropical high, which helps channel warm moisture plumes into midlatitude baroclinic systems developing near the East Asian coast or farther east and downstream waves. Conversely, the westward-moving CEP Clusters (b)-(c) are associated with significant skill increase for US and European forecasts after Day 6 (Figure 4f-g). We speculate this increase reflects the high inherent predictability of the stable large-scale circulation, rather than direct TC impacts. Although the exact physical mechanisms driving these bidirectional outcomes are not entirely clear, these findings highlight the complex relationship between the extratropical predictability and west-ward moving storms, which warrant future research.

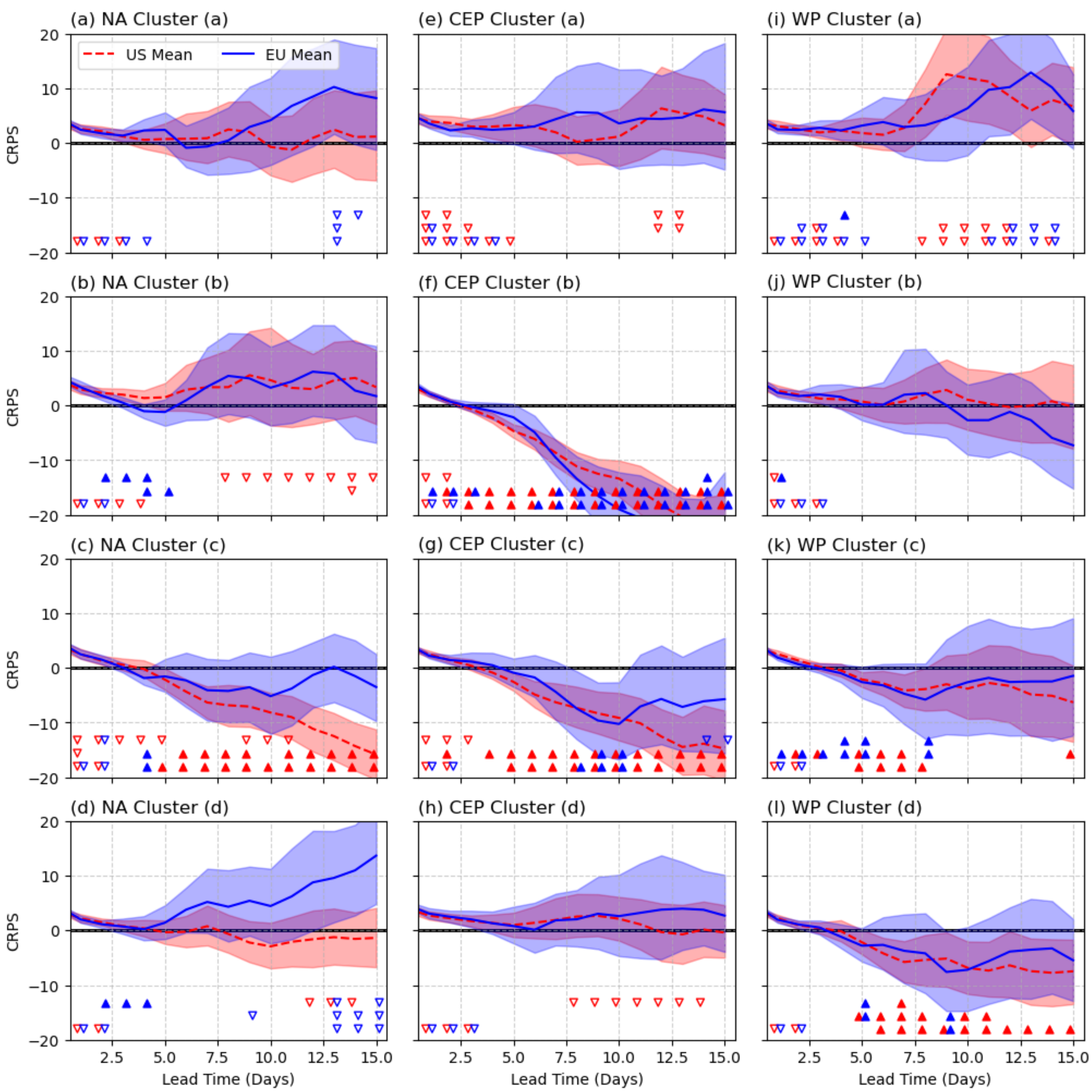

*Figure 4: NGCM skill differences between TC genesis forecasts (359 cases) and the genesis-free baseline (July-November of 2020-2022). Forecast skill is calculated for Z500 of Europe (20°W–42.5°E, 35–75°N; blue) and the US (65–128°W, 20–60°N; red). Shading indicates 95% bootstrap*

*confidence intervals estimated with 1000 iterations. Vertically stacked markers indicate lead times with statistically significant differences relative to the baseline (▲: significant increase; ▽: significant decrease) for ACCs, RMSE, and fair CRPS, from top to bottom.*

## 4. Summary and Discussion

This study provides a systematic analysis using objective track clustering and comparative model forecasts to assess the link between TC genesis and Northern Hemisphere extratropical forecasts out to 15 days. Our key conclusions are as follows:

1. The physical IFS and the AI-hybrid NGCM make comparable predictions of the extratropical 500-hPa geopotential height (Z500) after TC genesis. Case studies show qualitatively similar large-scale error growth patterns. When 108 out-of-sample cases are collectively evaluated, the two models show broadly comparable skill in the extratropics in Week 2, even though the NGCM exhibits a skill deficit in the tropics.
2. The statistical association between TCs and extratropical predictability is highly sensitive to the specific track archetype. While specific North Atlantic (NA), Central-Eastern Pacific (CEP), and Western Pacific (WP) clusters are associated with downstream skill degradation over the US and/or the European domains, certain westward-moving clusters (e.g., CEP Clusters (b)-(c)) are paradoxically related to statistically significant improvements in Week-2 skill.

Although the comparable performance of recent AI and physical models in forecasting the large-scale field (e.g., Z500) has been demonstrated, it is remarkable that this consistency holds during complex TC-midlatitude flow interactions. Given that the NGCM does not explicitly represent convective processes, its ability to match the IFS's downstream large-scale error growth in out-of-sample forecasts suggests that predicting the bulk upscale effects of TC convection is sufficient to capture the subsequent midlatitude wave response. Since the supplementary analysis (Figures S11–S13) suggests the stochastic NGCM greatly outperforms the deterministic NGCM in Week-2 extratropical forecasts, the comparable extratropical performance of physical and AI-hybrid models is likely model-specific and should not be generalized to other AI models without careful evaluation.

Ultimately, this work underscores the profound and multifaceted associations between TC track characteristics and extratropical weather predictability. While the newly developed NGCM shows certain limitations (e.g., tropics), this study highlights its potential as a valuable tool for studying the large-scale flow predictability. A model intercomparison project that evaluate more physical and AI models—particularly operational, purely data-driven models (e.g., Lang et al. 2026)—will help verify the robustness of the findings and offer insights into extratropical weather predictability.

## Conflict of Interest

G.Z. holds stocks of Alphabet, which is the parent company of Google Research, the developer of NeuralGCM. G.Z. declares no other conflicts of interest for this manuscript.

## Acknowledgement

G.Z. thanks Drs. Janni Yuval and Linus Magnusson for stimulating discussions about AI models, Drs. Christian M. Grams and Ron McTaggart-Cowan for insightful, constructive comments, and Ethan Cai for testing the track clustering algorithm. The research is supported by the U.S. National Science Foundation awards AGS-2327959 and RISE-253055.

## Data Availability Statement

The NeuralGCM model and code are accessible at https://neuralgcm.readthedocs.io/en/latest/. The ERA5 data is available at: https://console.cloud.google.com/marketplace/product/bigquery-public-data/arco-era5. The TIGGE data is archived by ECMWF and available at: https://apps.ecmwf.int/datasets/data/tigge/levtype=sfc/type=cf/. The TRACLUS package used by this study is available at: https://github.com/AdrielAmoguis/TRACLUS. The sample simulation code and the analysis code are available via Zenodo (Zhang, 2026).

Supplementary Materials

for

# Track-Dependent Links between Tropical Cyclones and Extratropical Predictability in Physical and AI Models

**Gan Zhang[1*]**

[1] Department of Climate, Meteorology & Atmospheric Sciences, University of Illinois Urbana-Champaign, Urbana, Illinois

*Corresponding Author: Gan Zhang (gzhang13@illinois.edu)

File Content

- Supplementary Discussion
- Figures S1–S13
- Table S1

## Parameter Choice of K-means Clustering

We conducted sensitivity tests of the k-means track clustering using a range of k values. Figure S2 shows the results of k = 3, 4, 5, and 6 for the TC tracks of the three examined basins. The results suggest the algorithm clusters tracks based on genesis locations and TC tracks. When k value increases, track clusters show more granular regional splitting. The results indicate that k=4 provides a meaningful separation into geographically distinct archetypes relevant to potential TC-midlatitude interactions. Furthermore, k = 4 prevents over small cluster sample sizes that makes it challenging to conduct inter-cluster comparison of forecast skills. We also note that previous studies of the North Atlantic TC tracks also use four clusters (e.g., Camargo et al., 2007, 2008; Kossin et al., 2010), which produces some similar track clusters. We adopt the silhouette score to evaluate the quality of K-means clustering (Figure S3). The silhouette score measures how similar an object is to its own cluster (cohesion) compared to others (separation). Ranging from -1 to +1, a higher value indicates better-defined clusters. Silhouette scores suggest k = 4 leads to a good balance between achieving relatively high values of the mean silhouette scores and keeping low values of individual silhouette scores relatively few. More details of clustering are available in the code repository (Data Availability). Lastly, we note that TCs in these track clusters also show differences in the genesis time (Figure S4) and lifespan (Figure S5), suggesting the spatial-temporal characteristics of TCs are linked.

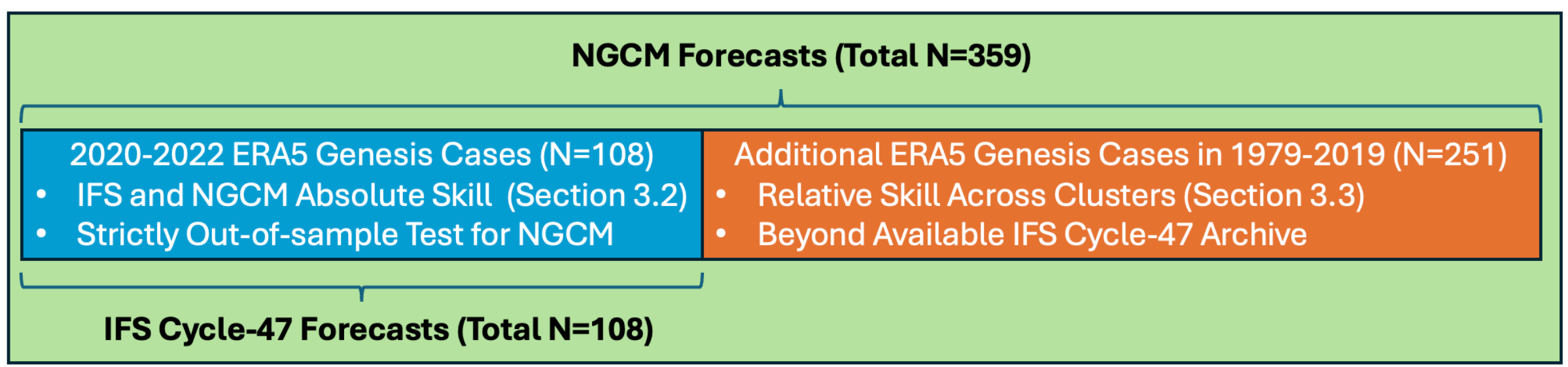

*Figure S1: Schematic of the experimental workflow and forecast case selection. Tracks (1979–2022) derived from ERA5 are clustered. A direct out-of-sample comparison with the operational IFS is performed using 108 cases from 2020–2022 (blue), while an extended sensitivity analysis by track cluster leverages 359 total NGCM forecasts, including 257 additional historical cases (orange). The date information of the forecast cases are available in Table S1.*

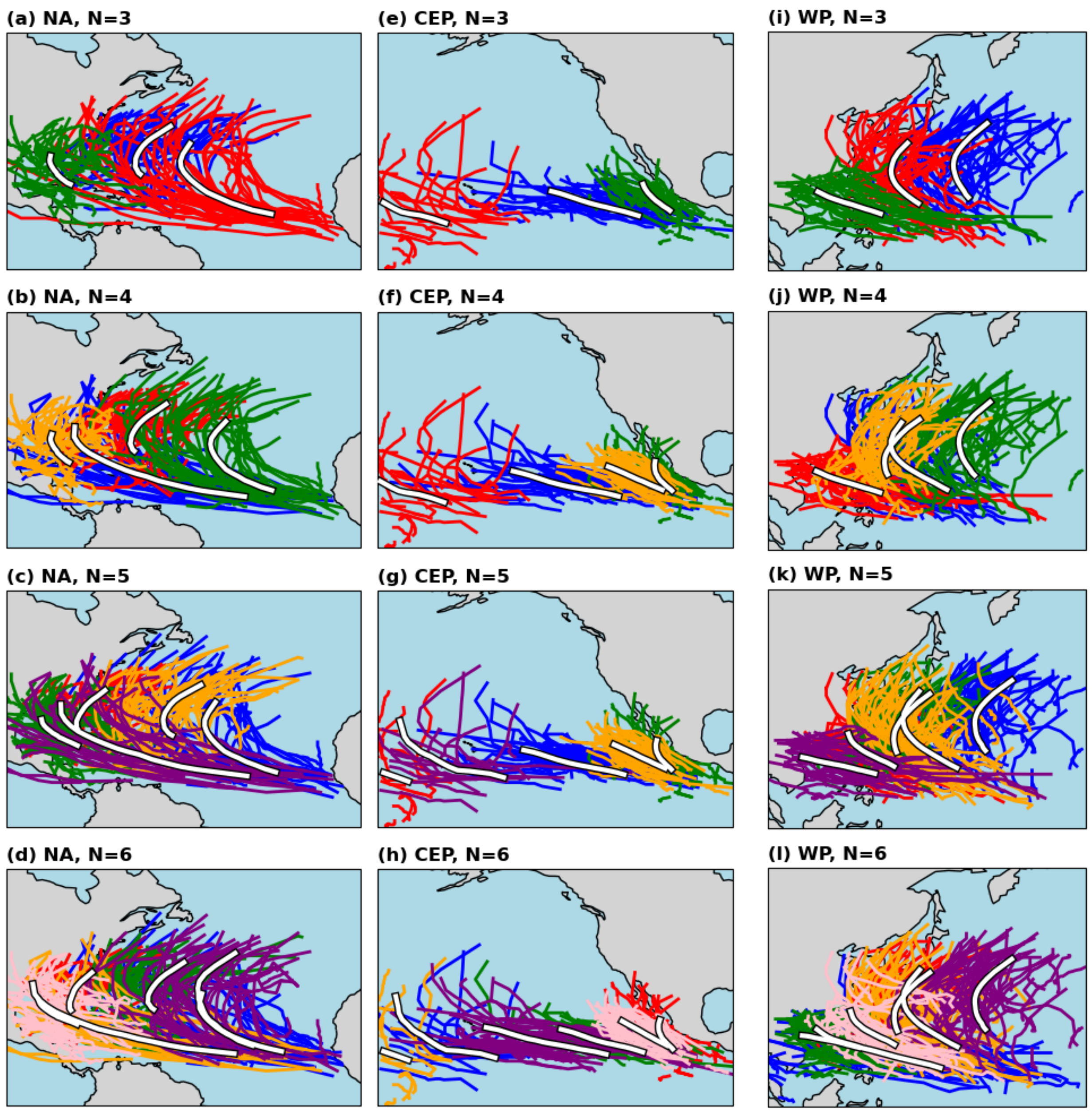

*Figure S2: Comparison of clustered historical TC tracks for varying cluster counts (N). Columns display (a-d) North Atlantic (NA), (e-h) Central Eastern Pacific (CEP), and (i-l) Western Pacific (WP) basins. Rows show varying cluster counts ($N \in \{3,4,5,6\}$). Individual tracks within each panel are color-coded according to cluster labels, and thick white lines represent cluster centroids (mean paths).*

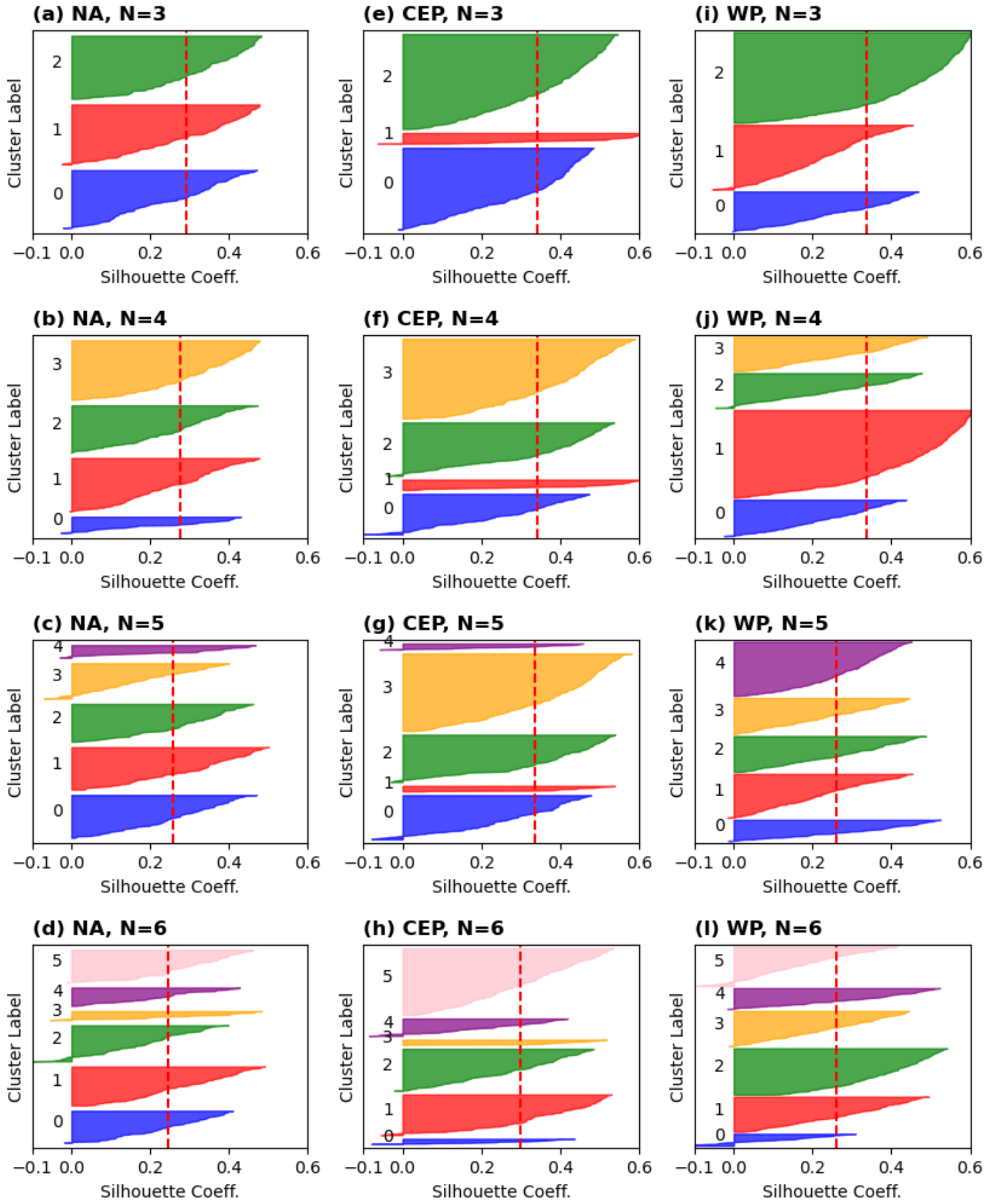

*Figure S3: Silhouette analysis of K-means clustering across varying cluster counts. Columns display (a-d) North Atlantic (NA), (e-h) Central Eastern Pacific (CEP), and (i-l) Western Pacific (WP) basins. Rows show varying cluster counts ($N \in \{3,4,5,6\}$). Each plot displays silhouette coefficients with a vertical dashed red line indicating the average score. Clusters are color-coded consistently for comparison.*

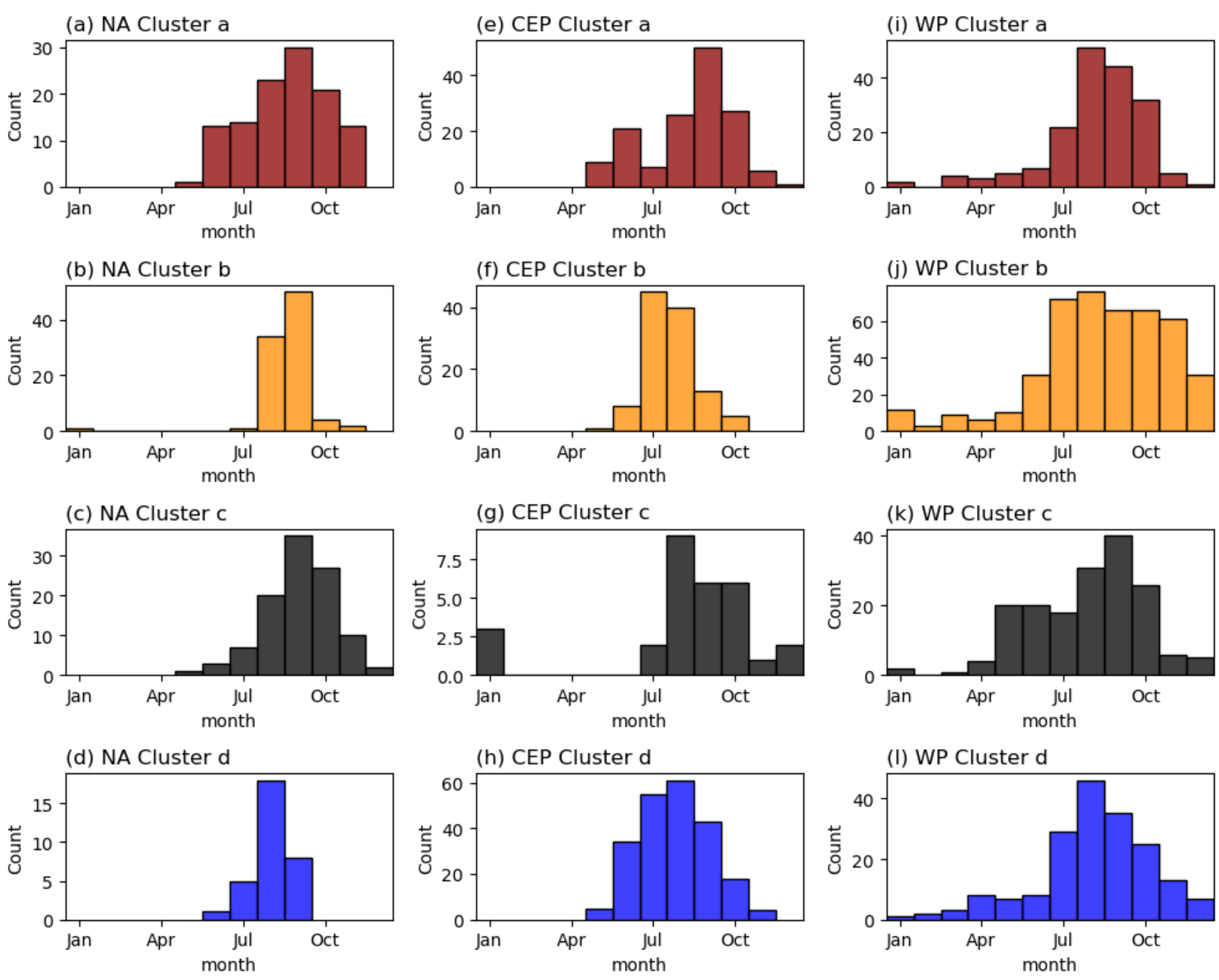

*Figure S4: Monthly distributions of genesis time for each TC cluster in (a-d) the North Atlantic (NA), (e-h) Central and Eastern Pacific (CEP), and (i-l) Western Pacific (WP) basins.*

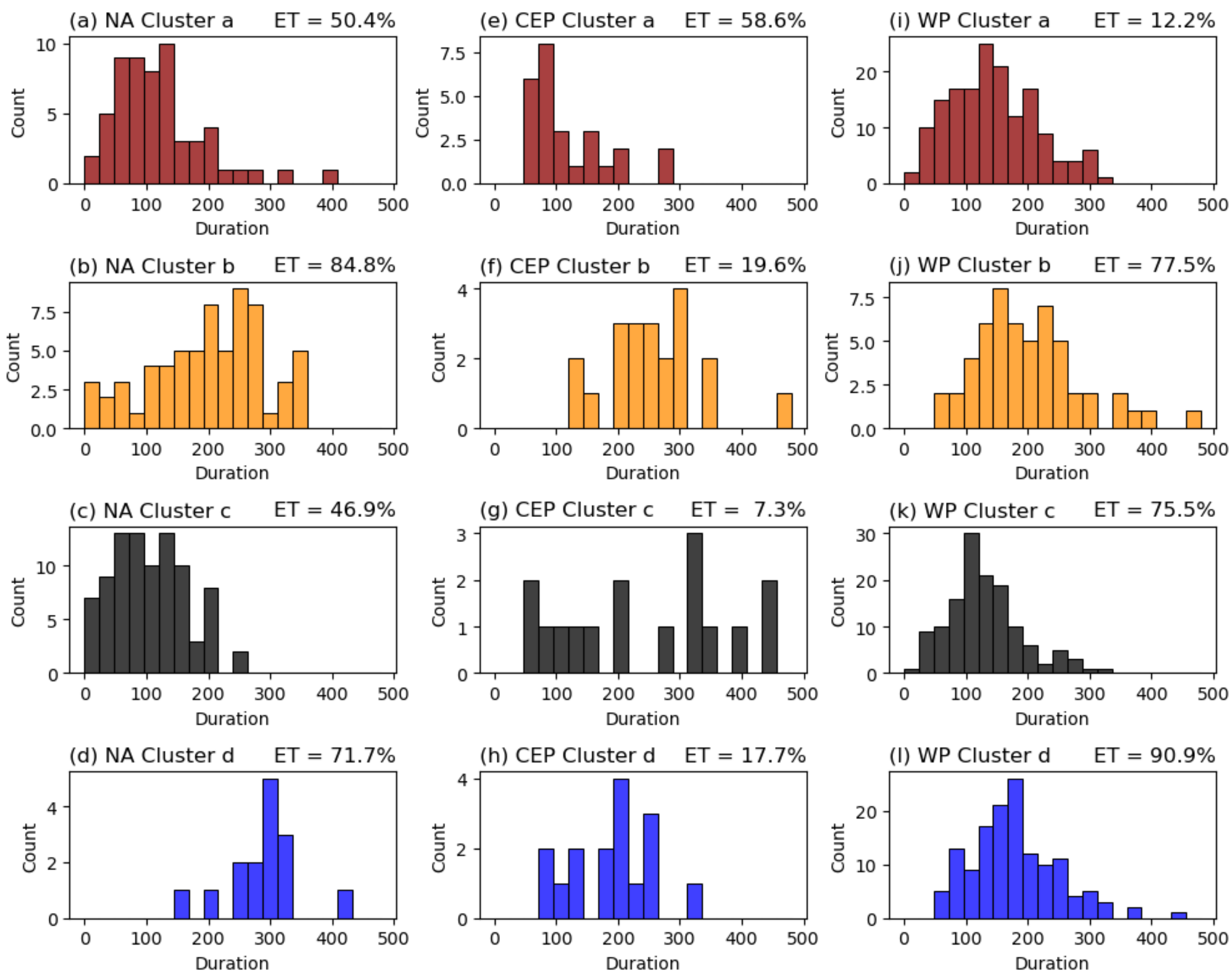

*Figure S5: Duration (hour) distributions for each TC cluster in (a-d) the North Atlantic (NA), (e-h) Central and Eastern Pacific (CEP), and (i-l) Western Pacific (WP) basins. The ratio of the extratropical transition (ET) percentages is calculated for each cluster and denoted in the top right of each panel. The ET information is from Han & Ullrich (2025), which determined the extratropical transition using threshold checks of average deep -layer wind shear and maximum stratosphere relative humidity near storms.*

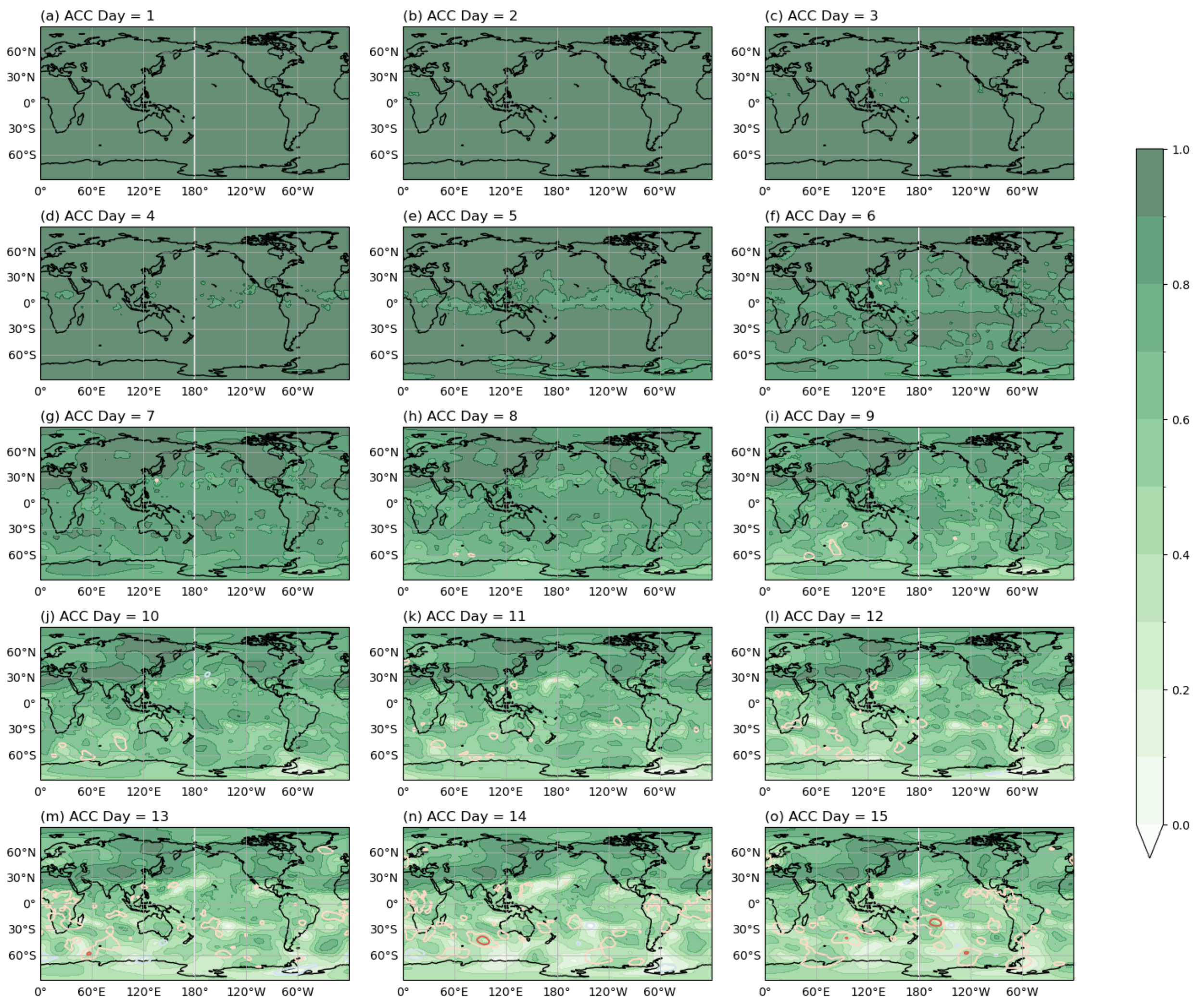

*Figure S6: Skill of IFS Z500 forecasts (ACC) for 108 cases during 2020-2022. Panels (a-o) show forecast skill for the lead time of Day 1 to Day 15, respectively. The forecasts are initialized around the genesis time (Methods). Shading represents the ACC values of a 20-member ensemble, while contours denote the difference between the 50-member ensemble and the 20-member ensemble. The contour level interval of ACC difference is 0.05, with zero contour line omitted for clarity. Smaller ACC values indicate less skillful predictions.*

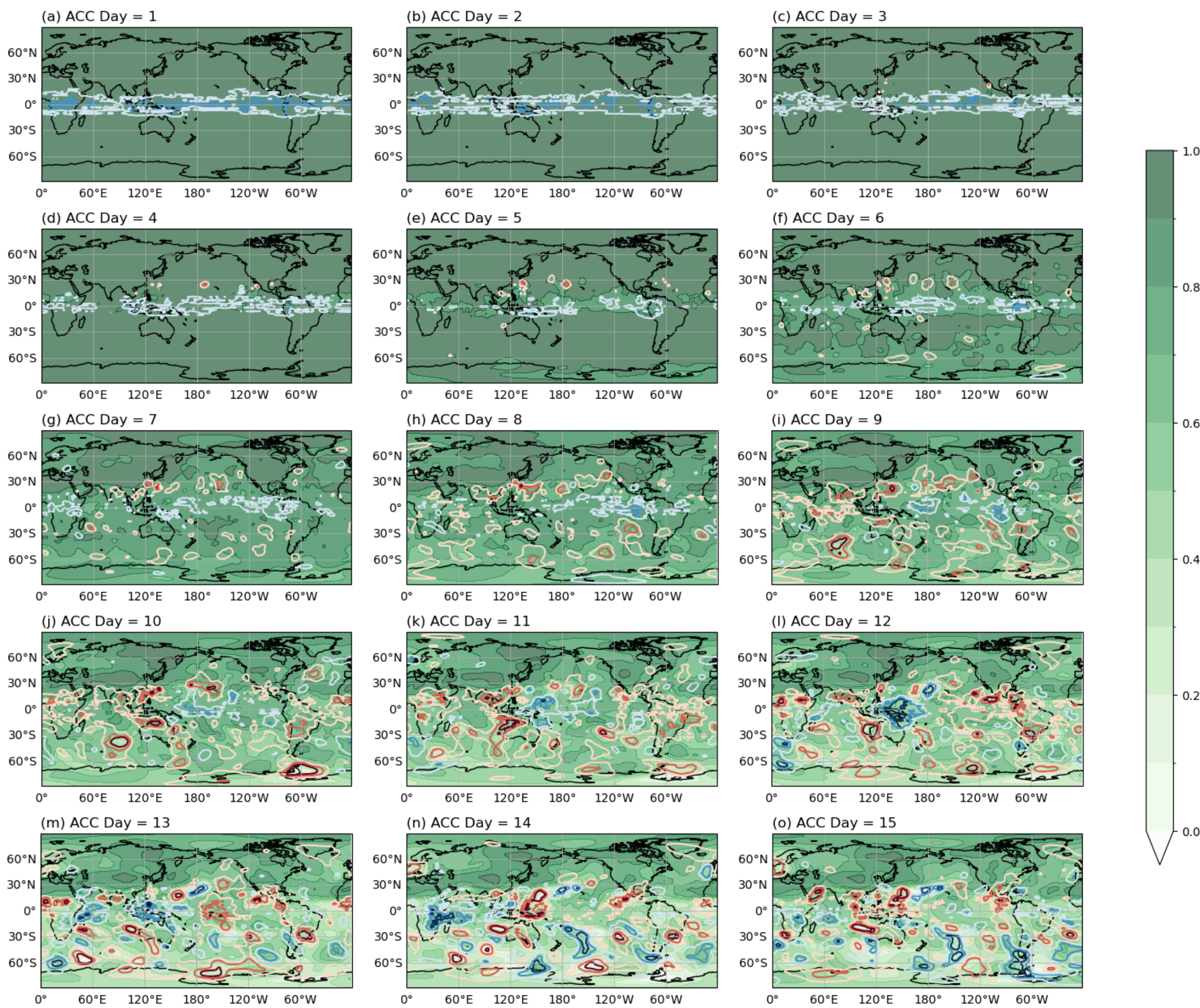

*Figure S7: Differences (contours) between the NGCM and the IFS skill of Z500 forecasts (ACC) for 108 cases during 2020-2022. The NGCM forecast skill is calculated using a 20-member ensemble of the stochastic NGCM. The other plotting settings are the same as Figure S4. Smaller ACC values indicate less skillful predictions.*

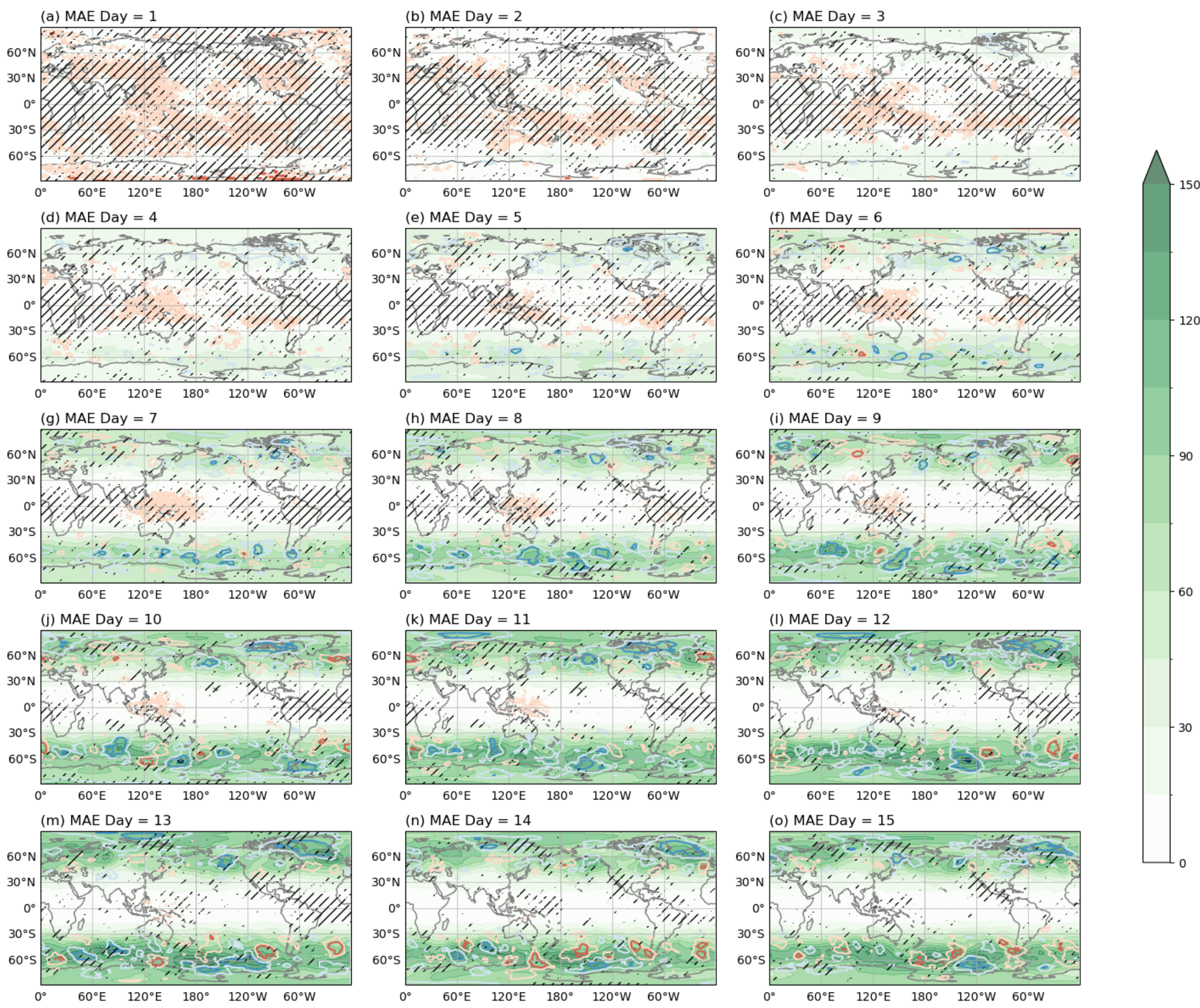

*Figure S8: Same as Figure S7, but for the mean absolute error (MAE). The contour levels of MAE differences are -20, -10, -5, 5, 10, 20 m. Hatching highlight the regions where IFS and NGCM are different at the 90%-confidence level, indicated by a sign test (DelSole & Tippett, 2016). Larger MAE values indicate less skillful predictions.*

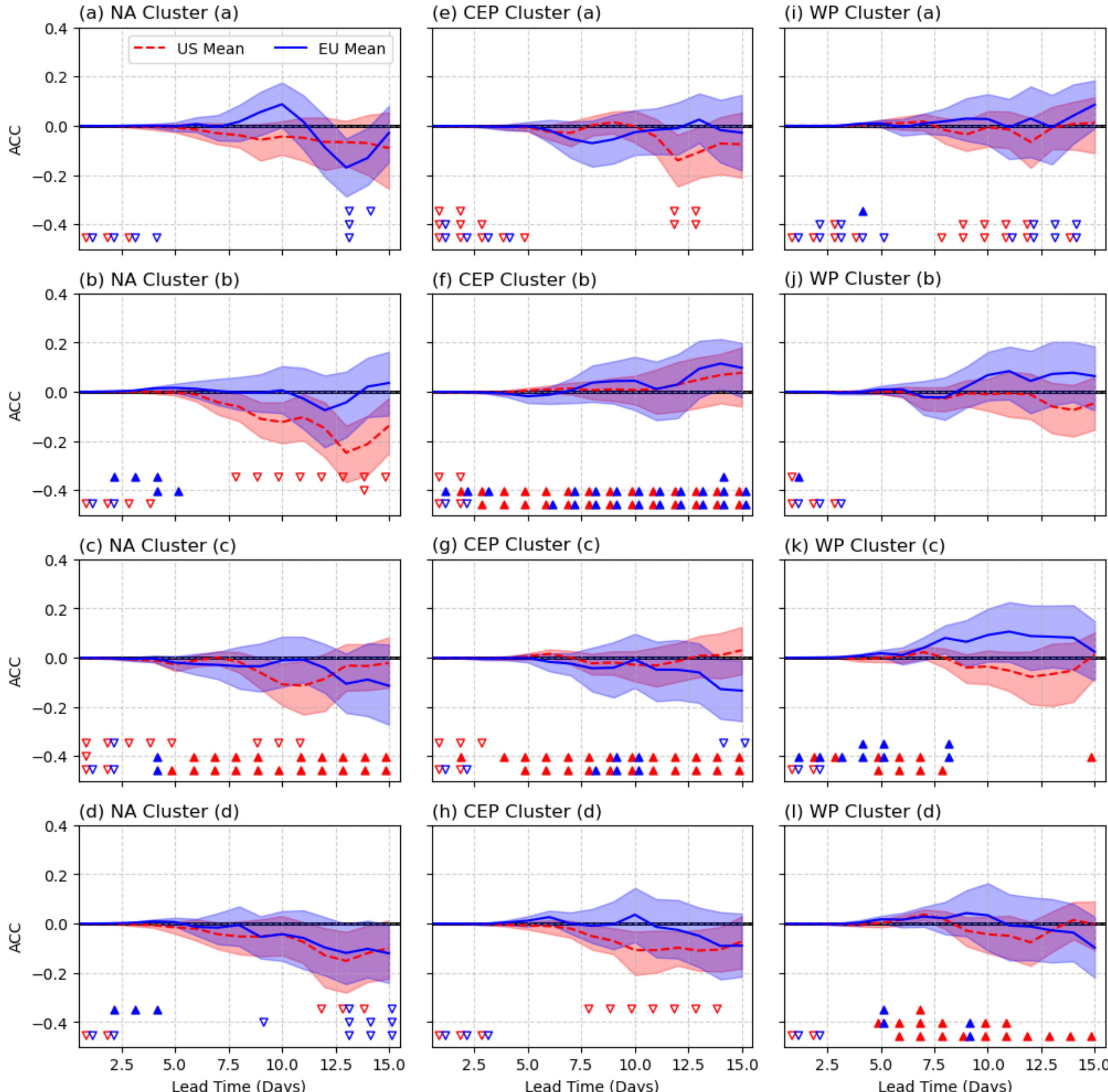

*Figure S9: Same as Figure 4, but for ACCs.*

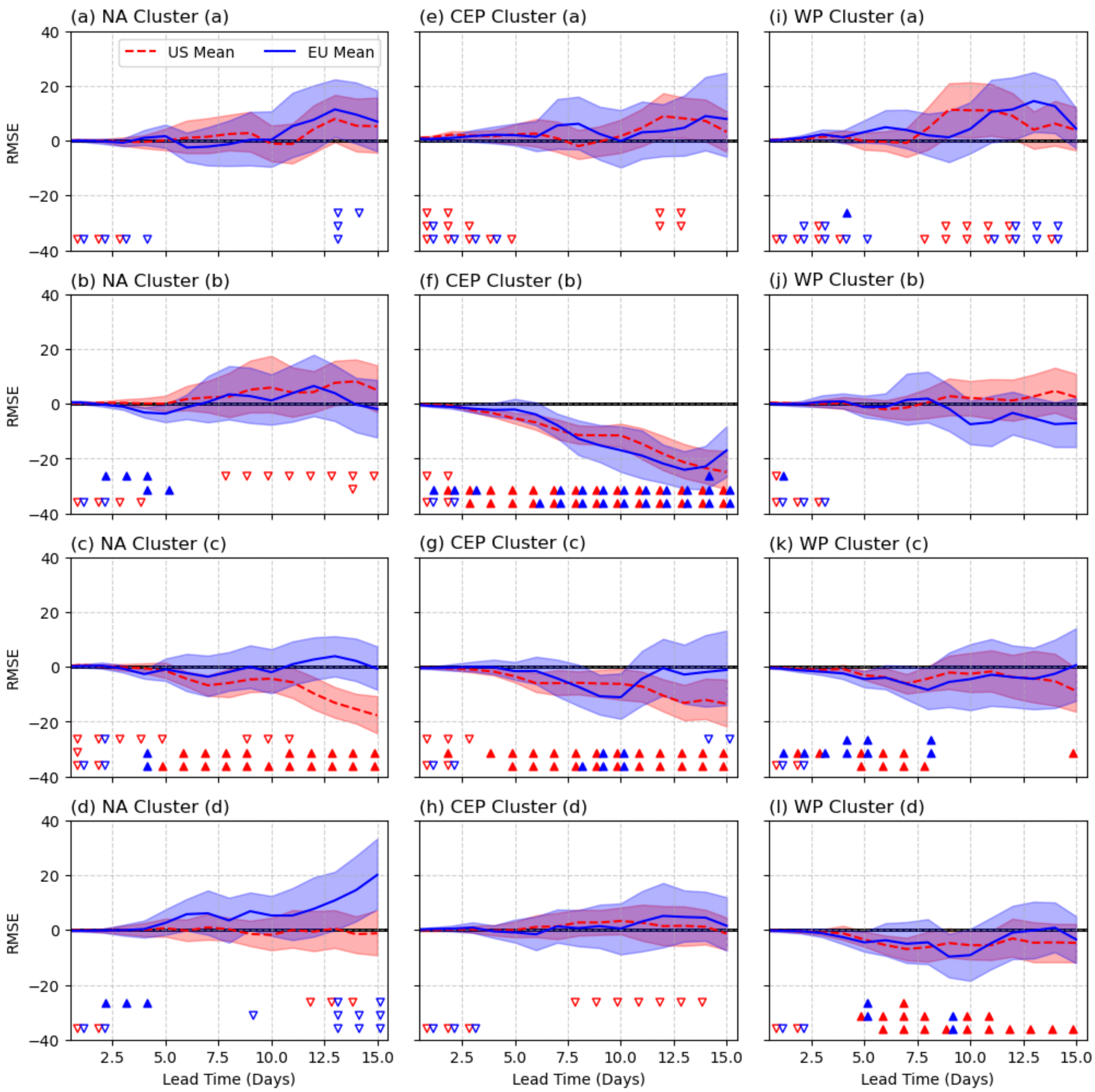

*Figure S10: Same as Figure 4, but for RMSE.*

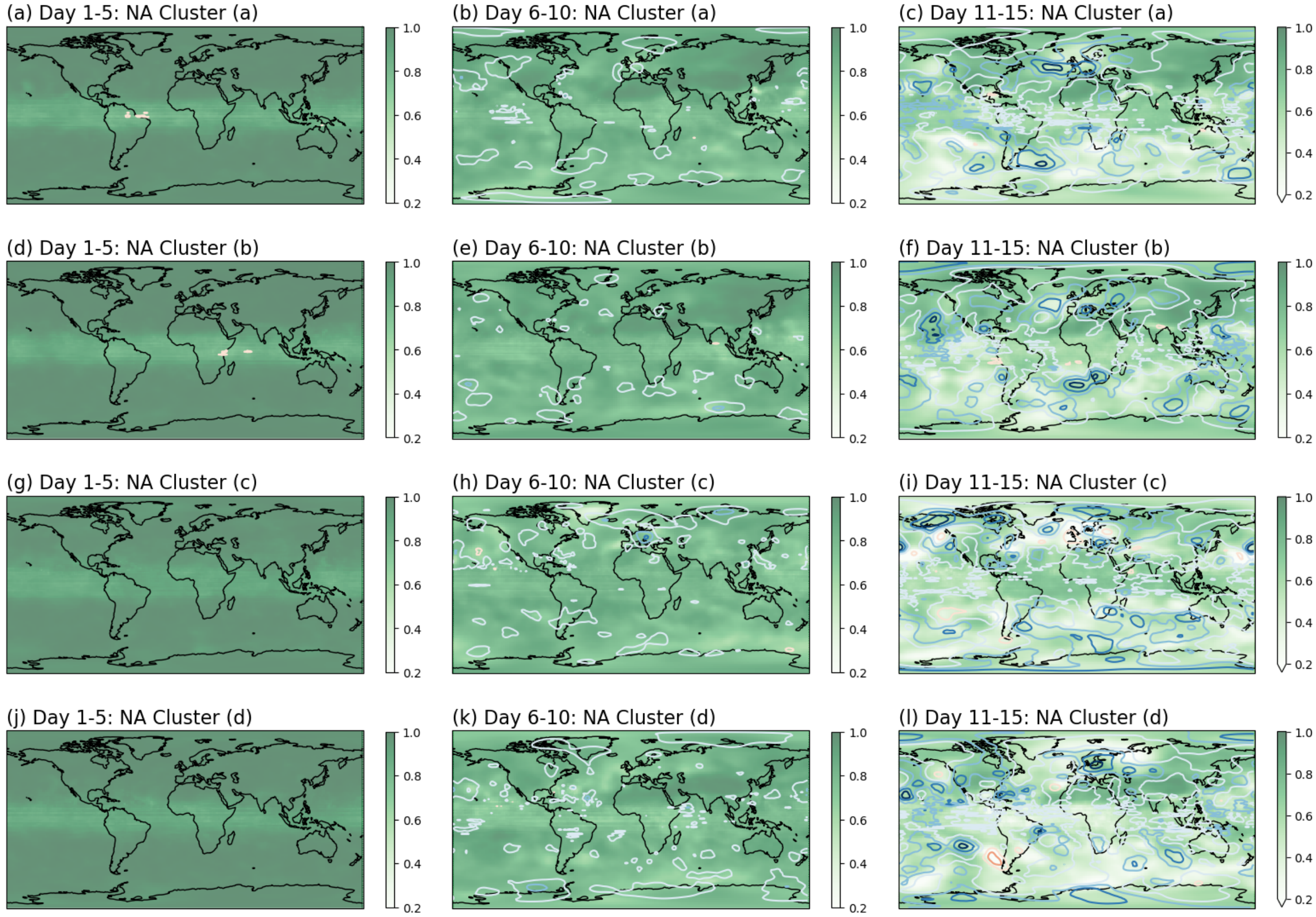

*Figure S11: Five-day mean skill of NGCM forecasts for each track cluster in the North Atlantic. Shading shows the ACC of 500-hPa geopotential height of the 20-member stochastic forecasts. Contours show the skill differences between the deterministic NGCM and the stochastic NGCM. The contour level interval of ACC difference is 0.1, with zero contour line omitted for clarity. (a-c) Mean ACC for the forecasts of track cluster a averaged over Day 1-5, 6-10, and 11-15, respectively. (d-f), (g-i), and (j-l): Same as (a-c), but for clusters b, c, and d. The calculations used the anomalies defined using the averages of the 108 cases during 2020-2022 for each forecast dataset.*

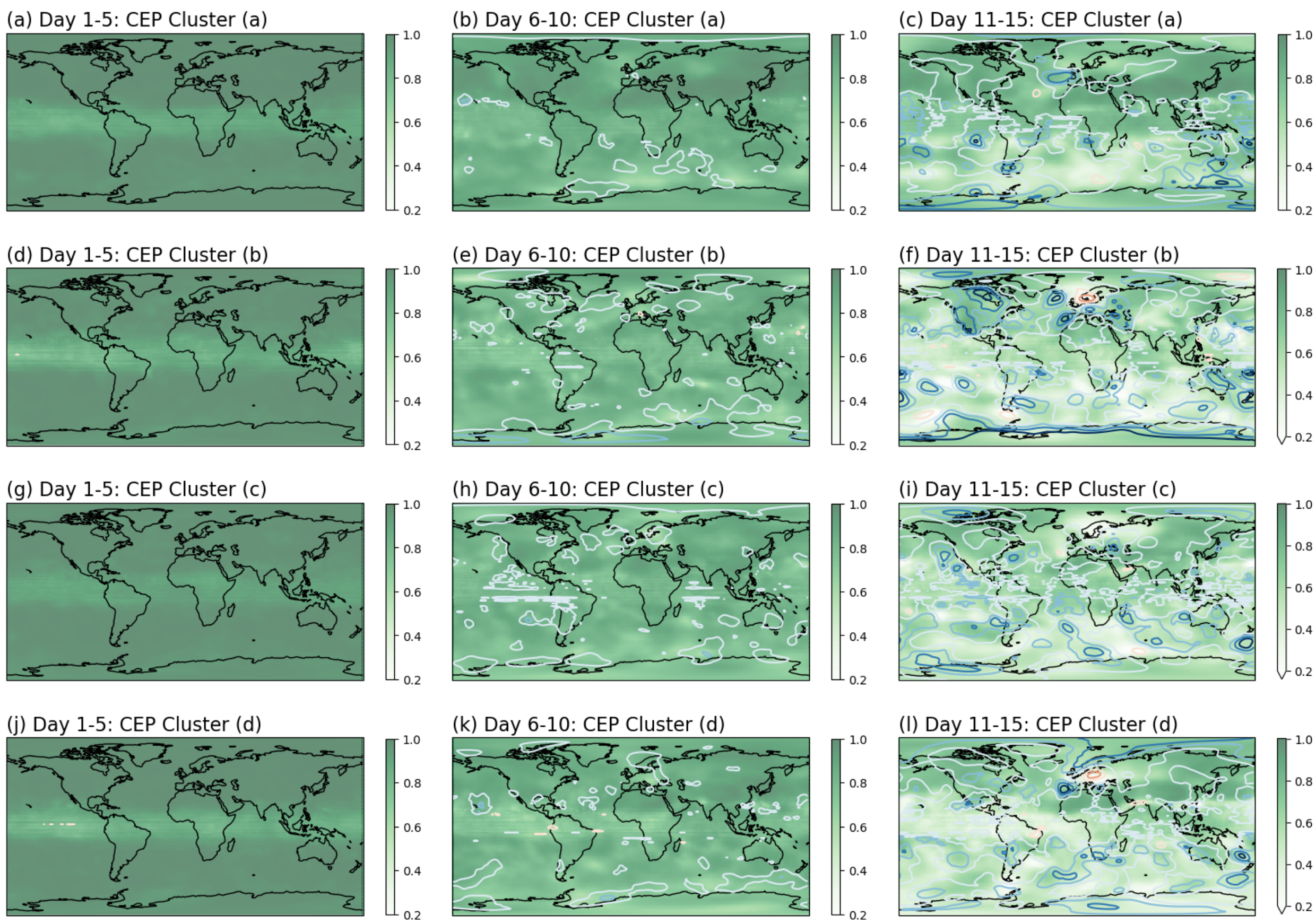

*Figure S12: Same as Figure S11, but for the Central and Eastern Pacific clusters.*

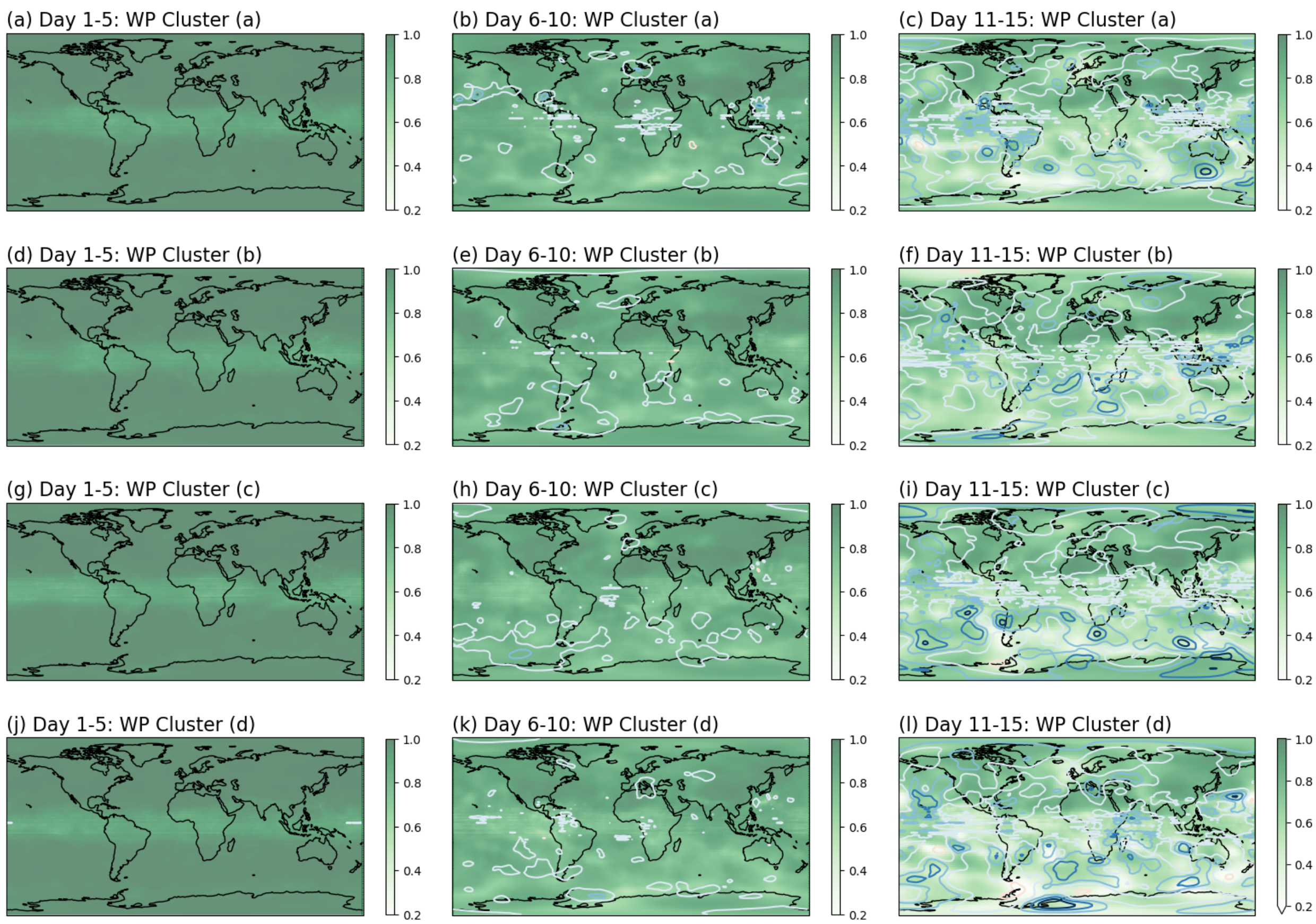

*Figure S13: Same as Figure S11, but for the Northwest Pacific track clusters.*

*Table S1 Information of the historical TC genesis cases. The dates correspond to the time of forecast initialization. The cases outside the NGCM training period are in italic with underlines (N=108) and analyzed in Section 3.2. All the cases (N=359) are analyzed in Section 3.3. Except for CEP Cluster (a), all the other clusters have at least 30 cases available during 1979-2022. The information is extracted using code in Zhang (2026).*

| WP Case # | Cluster (a) | Cluster (b) | Cluster (c) | Cluster (d) |
|---|---|---|---|---|
| 1 | 2019-11-19 | 2012-10-08 | 2016-09-02 | 2015-08-14 |
| 2 | 2019-11-24 | 2013-06-08 | 2016-09-27 | 2015-10-04 |
| 3 | 2019-12-21 | 2013-06-18 | 2017-07-21 | 2015-10-13 |
| 4 | 2020-05-09 | 2013-08-30 | 2017-09-10 | 2016-07-22 |
| 5 | *2020-09-15* | 2013-09-13 | 2017-10-18 | 2016-08-05 |
| 6 | *2020-10-11* | 2014-01-16 | 2018-07-20 | 2016-08-09 |
| 7 | *2020-10-20* | 2014-09-06 | 2018-07-17 | 2016-08-11 |
| 8 | *2020-10-24* | 2014-10-30 | 2018-08-06 | 2016-08-19 |
| 9 | *2020-10-28* | 2015-09-07 | 2018-08-15 | 2016-08-18 |
| 10 | *2020-10-29* | 2016-08-21 | 2018-08-18 | 2017-07-31 |
| 11 | *2020-11-08* | 2016-09-13 | 2018-08-16 | 2017-08-12 |
| 12 | *2020-11-09* | 2016-11-02 | 2018-09-20 | 2017-08-28 |
| 13 | 2021-02-16 | 2017-07-02 | 2018-09-27 | 2017-10-16 |
| 14 | 2021-05-29 | 2018-03-25 | 2018-11-19 | 2018-07-21 |
| 15 | 2021-06-11 | 2018-06-08 | 2019-02-19 | 2018-08-03 |
| 16 | *2021-07-18* | 2018-06-13 | 2019-07-13 | 2018-08-28 |
| 17 | *2021-07-17* | 2018-06-29 | 2019-08-04 | 2019-08-06 |
| 18 | *2021-08-05* | 2019-09-02 | 2019-08-01 | 2019-09-05 |
| 19 | *2021-09-06* | 2019-09-18 | 2019-09-27 | *2020-09-27* |
| 20 | *2021-09-22* | *2020-08-22* | 2019-10-05 | 2021-06-22 |
| 21 | *2021-10-02* | *2020-09-20* | 2019-10-18 | *2021-08-04* |
| 22 | *2021-10-09* | *2020-10-05* | 2019-11-03 | *2021-09-23* |
| 23 | 2021-12-13 | 2021-04-13 | 2019-11-10 | *2021-10-09* |
| 24 | 2022-04-09 | *2021-08-02* | *2020-08-01* | *2021-10-25* |
| 25 | 2022-06-29 | *2021-11-30* | *2020-08-28* | 2022-04-03 |
| 26 | *2022-08-21* | 2022-06-30 | *2020-08-31* | *2022-08-22* |
| 27 | *2022-09-24* | *2022-08-11* | *2021-09-06* | *2022-09-11* |
| 28 | *2022-10-14* | *2022-09-12* | *2022-07-30* | *2022-09-25* |
| 29 | *2022-10-28* | *2022-09-28* | *2022-08-28* | *2022-10-16* |
| 30 | *2022-10-28* | 2022-12-11 | *2022-09-09* | *2022-11-12* |
| | | | | |
| CEP Case # | Cluster (a) | Cluster (b) | Cluster (c) | Cluster (d) |
| 1 | 1982-11-22 | 2015-07-07 | 2017-07-08 | 2015-09-27 |
| 2 | 1984-09-03 | 2015-07-13 | 2017-07-22 | 2015-10-21 |
| 3 | 1987-09-22 | 2015-07-30 | 2017-08-12 | 2015-11-24 |
| 4 | 1988-08-28 | 2015-08-07 | 2017-08-18 | 2016-09-04 |
| 5 | 1992-01-27 | 2015-08-26 | 2017-09-11 | 2016-09-18 |
| 6 | 1992-09-08 | 2015-08-27 | 2017-09-14 | 2016-09-25 |
| 7 | 1993-08-11 | 2015-10-16 | 2018-06-06 | 2017-08-30 |
| 8 | 1994-08-05 | 2016-07-08 | 2018-06-28 | 2018-06-09 |
| 9 | 1994-08-22 | 2016-07-12 | 2018-06-30 | 2018-09-25 |
| 10 | 1994-10-22 | 2016-07-15 | 2018-08-07 | 2018-09-30 |
| 11 | 1995-01-03 | 2016-08-01 | 2018-08-05 | 2018-10-21 |
| 12 | 1997-07-23 | 2016-08-03 | 2018-09-09 | 2018-11-02 |
| 13 | 1997-09-02 | 2016-08-27 | 2019-09-02 | 2019-08-21 |
| 14 | 1997-12-01 | 2016-08-25 | *2020-07-07* | 2019-09-19 |
| 15 | 2002-08-28 | 2017-07-12 | *2020-08-09* | 2019-09-16 |
| 16 | 2002-10-28 | 2017-07-20 | *2020-09-12* | *2020-08-16* |
| 17 | 2006-08-20 | 2017-07-21 | *2020-09-20* | *2020-08-26* |
| 18 | 2009-10-18 | 2018-08-01 | *2020-09-29* | *2020-08-25* |
| 19 | 2013-08-15 | 2018-08-15 | *2020-11-03* | 2021-06-18 |
| 20 | 2014-10-13 | 2018-08-26 | 2021-05-31 | 2021-06-25 |
| 21 | 2015-07-07 | 2018-08-28 | *2021-07-17* | *2021-08-26* |
| 22 | 2015-08-21 | 2018-09-02 | *2021-07-31* | *2021-09-07* |
| 23 | 2015-08-22 | 2019-06-30 | *2021-08-07* | *2021-10-10* |
| 24 | 2015-09-25 | 2019-07-28 | *2021-08-22* | *2021-10-23* |
| 25 | 2015-10-01 | 2019-07-30 | 2022-06-15 | 2022-05-29 |
| 26 | 2015-12-28 | 2019-09-12 | 2022-06-20 | *2022-09-04* |
| 27 | 2016-01-06 | *2020-07-20* | *2022-07-15* | *2022-09-17* |
| 28 | 2016-10-03 | *2021-07-14* | *2022-07-26* | *2022-09-29* |
| 29 | 2018-09-30 | *2021-08-10* | *2022-08-07* | *2022-10-02* |
| 30 | | *2022-07-11* | *2022-09-01* | *2022-10-20* |

| NA Case # | Cluster (a) | Cluster (b) | Cluster (c) | Cluster (d) |
|---|---|---|---|---|
| 1 | 2013-06-05 | 2013-09-10 | 1980-08-03 | 2013-08-15 |
| 2 | 2013-09-13 | 2013-10-21 | 1985-09-16 | 2013-09-09 |
| 3 | 2013-10-03 | 2014-07-01 | 1988-09-10 | 2014-09-12 |
| 4 | 2015-06-14 | 2014-08-24 | 1989-09-11 | 2015-08-29 |
| 5 | 2016-08-03 | 2014-10-02 | 1996-07-06 | 2015-09-17 |
| 6 | 2016-09-13 | 2014-10-10 | 1996-07-25 | 2016-08-16 |
| 7 | 2016-11-22 | 2014-10-12 | 1996-08-22 | 2016-08-23 |
| 8 | 2017-06-20 | 2015-09-28 | 1996-08-26 | 2016-09-14 |
| 9 | 2017-08-08 | 2015-11-10 | 1998-08-23 | 2016-09-19 |
| 10 | 2017-08-23 | 2016-08-30 | 1998-09-14 | 2017-09-05 |
| 11 | 2017-09-06 | 2016-09-28 | 2001-08-30 | 2017-09-14 |
| 12 | 2017-10-05 | 2016-10-06 | 2003-09-06 | 2017-09-16 |
| 13 | 2018-05-27 | 2016-10-21 | 2004-08-27 | 2017-10-12 |
| 14 | 2018-09-04 | 2016-11-02 | 2004-09-02 | 2018-09-06 |
| 15 | 2019-07-11 | 2017-08-13 | 2005-07-10 | 2018-09-30 |
| 16 | 2020-06-02 | 2017-09-24 | 2007-08-13 | 2018-10-09 |
| 17 | *2020-07-23* | 2018-07-08 | 2008-07-03 | 2018-10-27 |
| 18 | *2020-09-13* | 2018-10-07 | 2008-09-01 | 2019-08-24 |
| 19 | *2020-09-18* | 2019-08-26 | 2010-08-30 | 2019-09-03 |
| 20 | *2020-10-03* | 2019-09-14 | 2012-08-01 | 2019-09-17 |
| 21 | *2020-10-05* | 2019-11-19 | 2012-08-20 | 2019-09-22 |
| 22 | *2020-10-25* | *2020-07-29* | 2015-08-16 | *2020-09-07* |
| 23 | *2020-11-01* | *2020-10-19* | 2015-08-24 | *2020-09-06* |
| 24 | *2020-11-14* | *2021-08-17* | 2017-08-31 | *2020-09-14* |
| 25 | *2021-08-15* | *2021-08-27* | 2018-08-30 | *2020-11-10* |
| 26 | *2021-09-12* | 2022-06-04 | 2018-09-04 | *2021-08-27* |
| 27 | *2022-09-24* | *2022-09-04* | *2020-08-20* | *2021-08-31* |
| 28 | *2022-10-07* | *2022-09-16* | 2021-06-30 | *2021-09-24* |
| 29 | *2022-11-02* | *2022-10-31* | *2021-08-14* | *2021-09-29* |
| 30 | *2022-11-07* | 2022-12-07 | *2022-07-01* | *2022-08-30* |